\documentclass[pra,twocolumn,superscriptaddress,longbibliography]{revtex4-1}
\usepackage{amssymb,amsmath,amsfonts,bm}
\usepackage{epsfig,graphicx}
\usepackage{times}
\usepackage{lipsum}
\usepackage{color}

\usepackage{hyperref}
\hypersetup{  colorlinks=true, linkcolor=blue, citecolor=red, urlcolor=blue  }

\usepackage{physics}

\begin{document}

\title{Leakage Suppression for Holonomic Quantum Gates}


\author{Bao-Jie Liu}

\affiliation{Department of Physics, Harbin Institute of Technology, Harbin 150001, China}
\affiliation{Institute for Quantum Science and Engineering, and Department of Physics,
Southern University of Science and Technology, Shenzhen 518055, China}

\author{Man-Hong Yung}  \email{yung@sustech.edu.cn}
\affiliation{Institute for Quantum Science and Engineering, and Department of Physics,
Southern University of Science and Technology, Shenzhen 518055, China}

\date{\today}

\begin{abstract}
Non-Abelian geometric phases acquired in cyclic quantum evolution can be utilized as natural resources for constructing robust holonomic gates for quantum information processing. Recently, an extensible holonomic quantum computation (HQC) was proposed and demonstrated in a recent superconducting experiment [T. Yan et al., Phys. Rev. Lett. 122, 080501 (2019)].  However, for the weakly anharmonic system, this HQC was given of low gate fidelity due to leakage to states outside of the computational subspace. Here, we propose a scheme that to construct nonadiabatic holonomic gates via dynamical invariant using resonant interaction of three-level superconducting quantum systems. Furthermore, the proposed scheme can be compatible with optimal control technology for maximizing the gate fidelity against leakage error. For benchmarking, we provide a thorough analysis on the performance of our scheme under experimental conditions, which shows that the gate error can be reduced by as much as 91.7\% compared with the conventional HQC.
Moreover, the leakage rates can be reduced to $10^{-3}$ level by numerically choosing suitable control parameter. Therefore, our scheme provides a promising way towards fault-tolerant quantum computation in a weakly anharmonic solid-state system.

\end{abstract}


\maketitle

\section{Introduction}
Quantum computation is considered to be a promising solution to some complex problems~\cite{Shor1997,Grover1997}, which will be very beneficial to many practical applications in quantum information processing~\cite{Nielsen2000,Cirac1995,Turchette1995}. However, due to inevitable noise and operational errors, constructing precise  quantum control is challenging in practice. Consequently, how to construct high-fidelity and noise-resistant quantum gates are of  fundamental importance to quantum computation. Geometric quantum computation (GQC)~\cite{Zanardi1999,gqc,b5,b6,Zhu2005,Berger2013,Chiara,Leek,Filipp} takes a different approach, which utilizes an unique property of quantum theory: the quantum state will acquire an Abelian or non-Abelian geometric phase after a cyclic evolution in non-degenerate or degenerate space.  An important property of geometric phase depends only on the overall properties of the evolutionary trajectory~\cite{Zhu2005,Berger2013,Chiara}. Therefore, quantum gates based on the geometric phases are robust to local control errors during a quantum evolution~\cite{Leek,Filipp}. Specifically, the geometric phase can be divided into a commutative real number, known as the adiabatic ``Berry phase''~\cite{b2} or nonadiabatic ``A-A phase''~\cite{AA}, and a non-commutative matrix (non-Abelian holonomy)~\cite{b3,Jones} which is the key ingredient in constructing holonomic quantum computation (HQC)~\cite{Zanardi1999}.

Previous applications of holonomic quantum computing (HQC)~\cite{Duan2001a,lian2005,b4} need to meet adiabatic condition to avoid transitions between different sets of eigenstates. However, adiabatic conditions imply lengthy quantum gate operation time. Therefore, adiabatic HQC is severely limited by decoherence caused by the environment noises~\cite{Liu2017PRA,Zhang2015Srep,Song2016NJP}. It was later discovered that if we construct a driving Hamiltonian with time-independent eigenstates and establish a holonomic gate, we can achieve nonadiabatic holonomic quantum computation (NHQC)~\cite{Sjoqvist2012,Xu2012,xue2017,Abdumalikov2013,Feng2013,Arroyo-Camejo2014,s1,Sekiguchi2017,Zhou2017,Egger2018,xu2018,Zhu2019,Han2020}. However, this condition imposes stringent requirements on the driving Hamiltonian; the systematic errors would introduce additional fluctuating phase shifts, smearing the geometric phases~\cite{Johansson,Zheng,jun2017}.
Recently, a flexible nonadiabatic holonomic quantum computation (NHQC+)~\cite{Liu2019} was proposed to avoid the constraints of NHQC and improve the robustness against systematic errors. However, these existing NHQC+ schemes~\cite{Liu2019,Yan2019,Sai2020,Kang2020} cannot solve the leakage error caused by the presence of noncomputational levels when driven by control pulses in many physical quantum systems such as superconducting qubits~\cite{Yan2019}.

Here, we shall come up with a scheme to construct nonadiabatic holonomic gates via dynamical invariant~\cite{LRI1} in a resonant three-level quantum system of superconducting qubit. In this way, our scheme maintains both flexibility and robustness against certain types of noises. Furthermore, our proposed scheme is compatible with optimal control technology to suppress the leakage error and thus maximize the gate fidelity. For benchmarking, we provide a thorough analysis on the performance of our scheme with recent experimental parameters; we found that the leakage rates can be limited below $10^{-3}$ by choosing proper optimal parameter, which is greatly improved as compared to the performance of conventional nonadiabatic HQC~\cite{xu2018,Yan2019}.

\section{The basic model}

\begin{figure}[htbp]
\centering
\includegraphics[width=8.5cm]{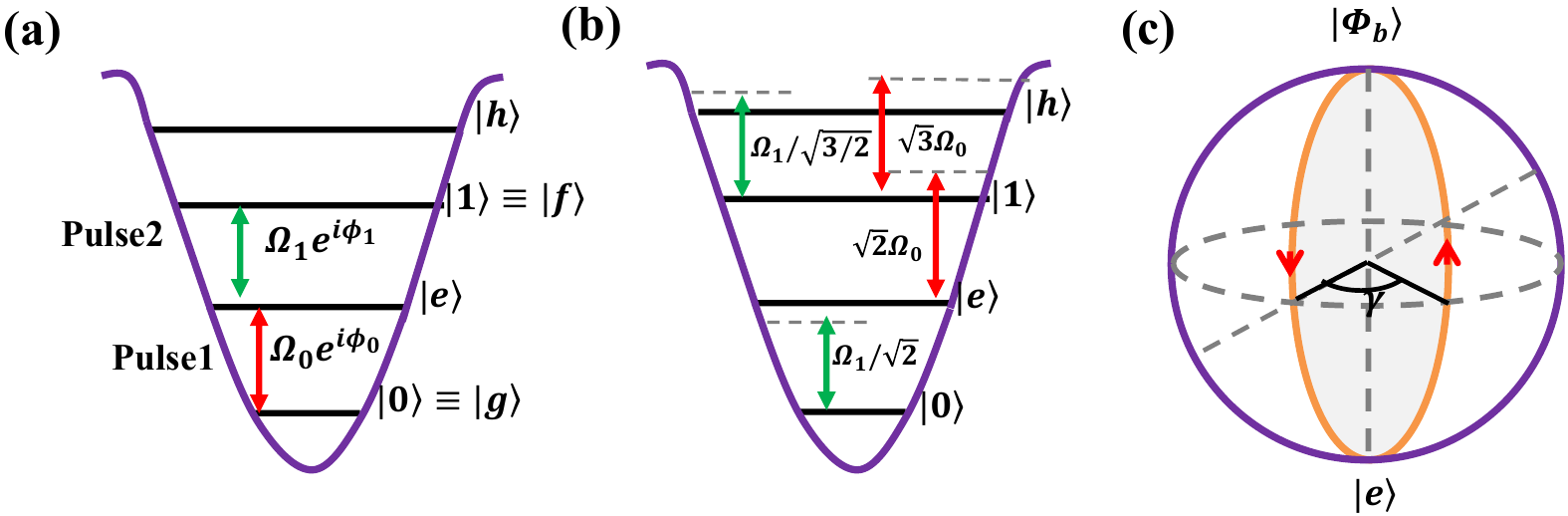}
\caption{Illustration of our proposed implementation. (a) HQC scheme using resonate three levels of an superconducting Xmon qubit as proposed in this work. Two pulses with Rabi frequencies of $\Omega_{0}(t)$ and $\Omega_{1}(t)$ are used. (b) Cross coupling and leakage to higher level in a many-level system with weak anharmonicity $\alpha$ in an Xmon type of superconducting device. For example, the Stokes pulse $\Omega_{1}$ that resonantly couples $|e\rangle$  and $|1\rangle$  in (a) can now also introduce an off-resonant coupling between $|0\rangle$  and $|e\rangle$ in an Xmon. (c) Conceptual explanation for holonomic quantum gate.}\label{setup}
\end{figure}

Here we consider a superconducting Xmon system demonstrated in Ref.~\cite{xu2018,Yan2019}, which has a weakly anharmonic potential. The lowest four energy levels of our Xmon qutrit are denoted by $|g\rangle$, $|e\rangle$, $|f\rangle$ and $|h\rangle$, as shown in Fig. \ref{setup}(a). Here, the ground state $|g\rangle\equiv|0\rangle$ and the second excited state $|f\rangle\equiv|1\rangle$ form our the computational subspace. And the first excited state  $|e\rangle$ and third excited state $|h\rangle$ denote non-computational state. The
system is resonantly driven by two microwave pulses to realize the transitions of $|0\rangle \leftrightarrow|e\rangle$ and $|e\rangle \leftrightarrow|1\rangle$. The system Hamiltonian in the basis $\{|0\rangle, |e\rangle, |1\rangle, |h\rangle\}$ is given by (hereafter $\hbar\equiv1$)
\begin{equation}\label{Original}
H=\left(\begin{array}{cccc}0 & f & 0 & 0 \\ f & \omega_{0} & \sqrt{2}f& 0 \\ 0 & \sqrt{2}f & \omega_{0}+\omega_{1} & \sqrt{3}f \\ 0 & 0 & \sqrt{3}f& \omega_{0}+\omega_{1}+\omega_{2}\end{array}\right) ,
\end{equation}
where the pulse field $f=\Omega_{0} \cos \left(\omega_{0} t+\phi_{0}\right)+\frac{\Omega_{1}}{\sqrt{2}} \cos \left(\omega_{1} t-\phi_{1}\right)$ with the driving amplitude $\Omega_{0,1}(t)$, frequency $\omega_{0,1}$, and phase $\phi_{0,1}(t)$. The corresponding transition energies are expressed by $\omega_{0,1,2}$, with the ground energy set to zero. And the intrinsic anharmonicity of the Xmon system is taken as $\alpha=\omega_{0}-\omega_{1}$. Here, using the rotating wave approximation and moving to the interaction frame, the Hamiltonian is
rewritten as
\begin{equation}\label{INter}
H_{I}=\frac{1}{2}\left(\begin{array}{cccc}0 & \Omega_{0} e^{i \phi_{0}} & 0 & 0 \\ \Omega_{0} e^{-i \phi_{0}} & 0 & \Omega_{1} e^{-i \phi_{1}} & 0 \\ 0 & \Omega_{1} e^{i \phi_{1}} & 0 & 0 \\ 0 & 0 & 0 & 0\end{array}\right)+H_{\text {leak}} \ .
\end{equation}
Here, $H_{\text {leak}}$ describes the cross coupling and leakage caused by higher excited energy [see Fig. \ref{setup}(b)], with the form
 \begin{small}
\begin{equation}
H_{\text {leak}}=\frac{\sqrt{2} }{2}\left(\begin{array}{cccc}0 & \frac{\Omega_{1}}{2} e^{-i\left(\phi_{1}+\alpha t\right)} & 0  & 0\\ \frac{\Omega_{1}}{2} e^{i(\phi_{1}+\alpha t)} & 0 & \Omega_{0} e^{i\left(\phi_{0}+\alpha t\right)} & 0 \\ 0 &  \Omega_{0} e^{-i\left(\phi_{0}+\alpha t\right)} & 0 & M \\ 0 & 0 & M^{*} & 0\end{array}\right).
\end{equation}
\end{small}
where $M\equiv\sqrt{3/2}\Omega_{0} e^{i\left(\phi_{0}+2 \alpha t\right)}+\frac{\sqrt{3}\Omega_{1}}{2} e^{-i(\phi_{1}-\alpha t)}$.
For the case $|\alpha| \gg |\Omega_{\mathrm{0}, \mathrm{1}}|$, the term $H_{\text {leak}}$ averages out to zero.

For our purpose, we assume that $\Omega_{0}(t)$ and $\Omega_{1}(t)$  have  the  same  time dependence, which means that the time-dependent driving amplitude can be parameterized as $\Omega_{0}(t)=\Omega(t)\sin{\frac{\theta}{2}}$, $\Omega_{1}(t)=\Omega(t)\cos{\frac{\theta}{2}}$, where the ratio of the two pulses $\Omega_{0}(t)/\Omega_{1}(t)$ to be time-independent, i.e., $\tan(\theta/2)=\Omega_{0}(t)/\Omega_{1}(t)$ is a constant. Besides, the time-independent relative phase is set as $\phi=\phi_{1}(t)-\phi_{0}(t)$.   Consequently, the effective Hamiltonian  can be given by
\begin{equation}\label{Eff} H_{e}(t)=\Omega(t)\cos\phi_{0}(t)\hat{T}_{1}+\Omega(t)\sin\phi_{0}(t)\hat{T}_{2}.
\end{equation}
where $\hat{T}_{1} \equiv (|\Phi_{b}\rangle\langle e|+H.c.)/2$, $\hat{T}_{2}\equiv (i|\Phi_{b}\rangle\langle e|+H.c.)/2$, and $\hat{T}_{3}\equiv (|\Phi_{b}\rangle\langle\Phi_{b}|-|e\rangle\langle e|)/2$
are the $SU(2)$ algebra of unitary $2\times2$ matrices, obeying the commutation relation $[\hat{T}_{a},\hat{T}_{b}]=i\varepsilon^{abc}\hat{T}_{c}$. And the time-independent bright state $|\Phi_{b}\rangle \equiv\sin(\theta/2)|0\rangle+\cos(\theta/2)e^{i\phi}|1\rangle$
is orthogonal to the excited state $|e\rangle$. Note that dark state $|\Phi_{d}\rangle =\cos(\theta/2)|0\rangle-\sin(\theta/2)e^{i\phi}|1\rangle$ is now decoupled from states $|\Phi_{b}\rangle$ and $|e\rangle$.

For the $SU(2)$ dynamical symmetry Hamiltonian $H_{e}(t)$ in Eq. (\ref{Eff}), the dynamical invariant $I(t)$, such that  $\frac{{dI}}{{dt}} \equiv \frac{{\partial I}}{{\partial t}} + \frac{1}{{i\hbar }}\left[ {I,H} \right] = 0$, is given by~\cite{LRI1}
\begin{equation}\label{IIII}
\begin{aligned} I(t) &=\frac{G{_{0}}}{2}\left(\cos \eta \sin \chi \hat{T}_{1}+\sin \eta \sin\chi \hat{T}_{2}+\cos \chi \hat{T}_{3}\right) \\ &= \frac{G{_{0}}}{2}\left(\begin{array}{cc}\cos \chi & \sin\chi e^{-i\eta} \\ \sin\chi e^{i\eta} &  -\cos \chi \end{array}\right) \ , \end{aligned}
\end{equation}
where $G_{0}$ is an arbitrary constant to make $I (t)$ dimensionless. To be explicit, we rewrite the dynamical invariant $I(t)$ in basis $\{|0\rangle,|e\rangle,|1\rangle\}$ as
\begin{equation}\label{Ithree}
    I(t)=\left[\begin{array}{ccc}
-c_{\chi} s_{\theta / 2}^{2} & s_{\chi} s_{\theta / 2} e^{-i \eta} & \frac{1}{2} c_{\chi} s_{-\theta} e^{-i \phi} \\
s_{\chi} s_{\theta / 2} e^{i \eta} & c_{\chi} & s_{\chi} c_{\theta / 2} e^{-i(\eta-\phi)} \\
\frac{1}{2} c_{\chi} s_{-\theta} e^{i \phi} & s_{\chi} c_{\theta / 2} e^{i(\eta-\phi)} & -c_{\chi} c_{\theta / 2}^{2}
\end{array}\right] \ ,
\end{equation}
where ${c_x} \equiv \cos x$ and ${s_x} \equiv \sin x$. And the time-dependent auxiliary parameters $\chi$ and $\eta$ satisfy the differential equations,
\begin{equation}\label{diffequ}
\Omega(t)=\frac{\dot{\chi}}{\sin \left(\phi_{0}-\eta\right)}, \quad \phi_{0}(t)=\eta-\arctan \left(\frac{\dot{\chi}}{\dot{\eta}\tan\chi} \right).
\end{equation}
The eigenvectors of the invariant are given by $\left|\mu_{0}(t)\right\rangle=\left(\cos \frac{\chi}{2} \mathrm{e}^{-\mathrm{i} \frac{\eta} {2}},\sin\frac{\chi}{2}  \mathrm{e}^{\mathrm{i} \frac{\eta} {2}}\right)^{\text{T}}$ and $\left|\mu_{1}(t)\right\rangle=\left(\sin \frac{\chi}{2} \mathrm{e}^{-\mathrm{i} \frac{\eta} {2}},-\cos\frac{\chi}{2}  \mathrm{e}^{\mathrm{i} \frac{\eta} {2}}\right)^{\text{T}}$.

\begin{figure}[htbp]
\centering
\includegraphics[width=8.7cm]{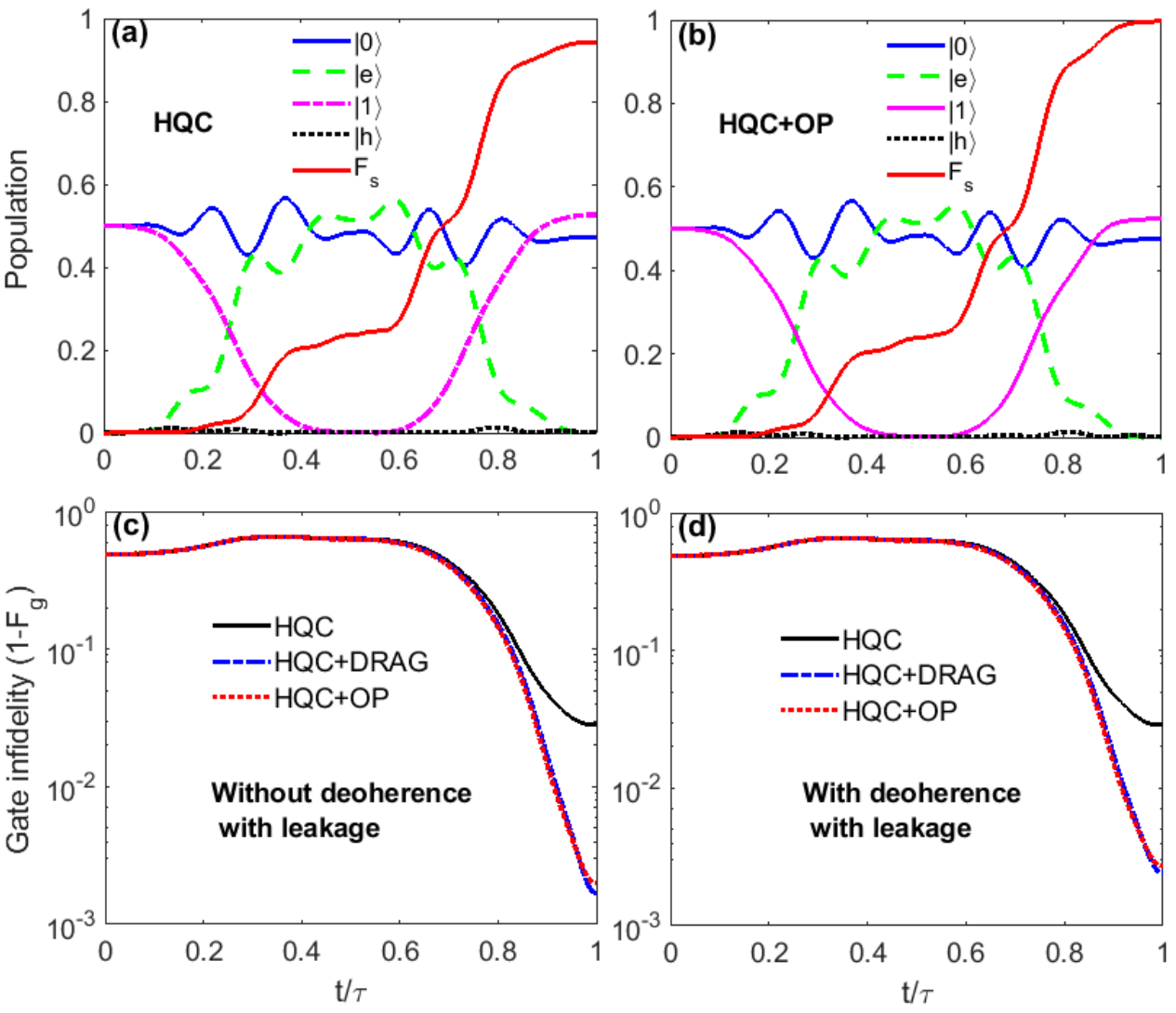}
\caption{ Illustration of the performance of the proposed holonomic Z gate. State population and fidelity dynamics: (a) Z gates of HQC (without optimization) and (b) HQC+OP (with optimization), as a function of $t/\tau$ with the initial state being $|\psi(0)\rangle=(|0\rangle+|1\rangle)/\sqrt{2}$. The dynamics of the Z gate infidelities ($1-F_{g}$) of HQC, HQC+DRAG, and HQC+OP with different settings of decoherence and leakage parameters in (c) and (d).}\label{Zgate}
\end{figure}

\begin{figure}[htbp]
\centering
\includegraphics[width=8.7cm]{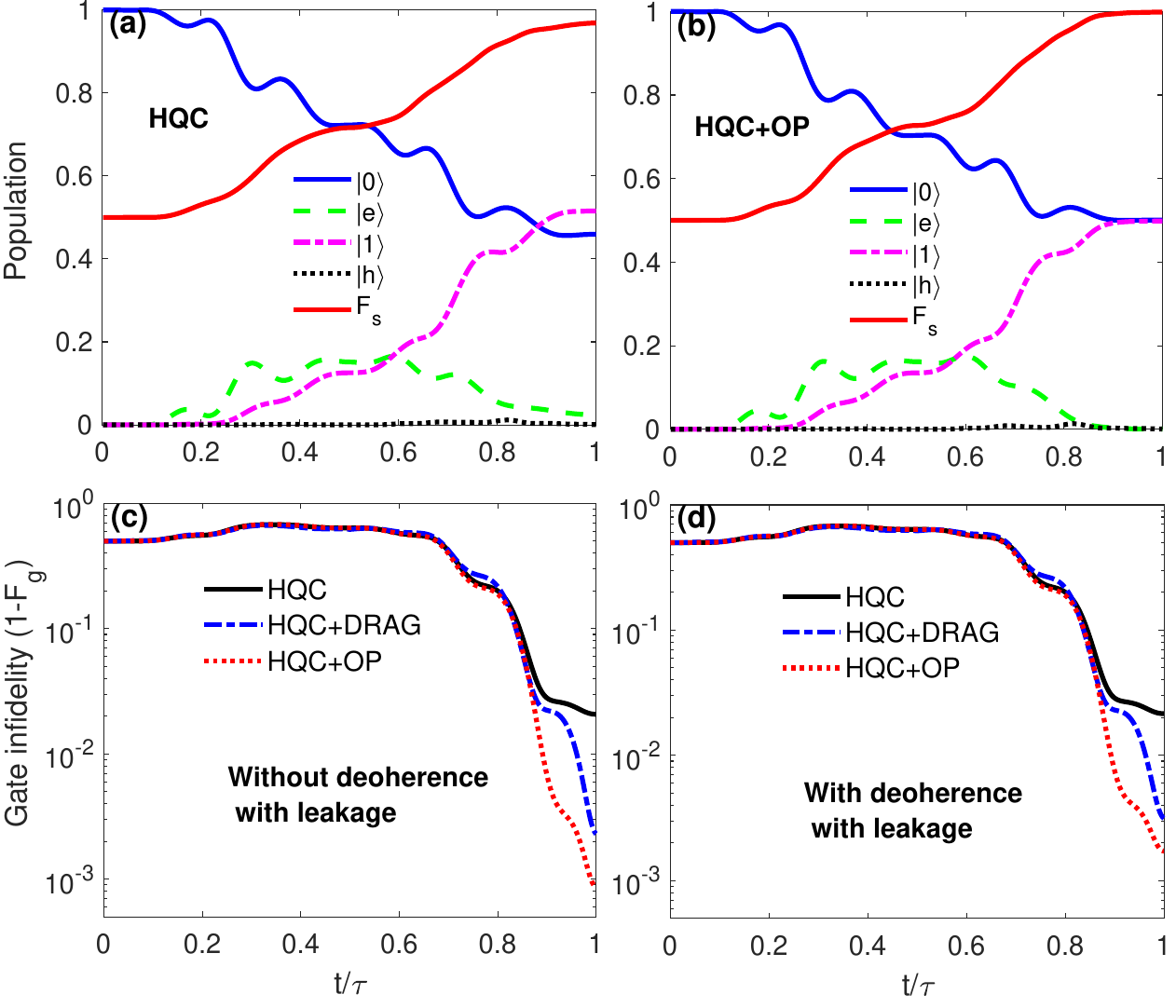}
\caption{ State population and fidelity dynamics: (a) Hardmard gates of HQC and (b) HQC+OP, as a function of $t/\tau$ with the initial state being $|\psi(0)\rangle=|0\rangle$. The dynamics of the Hardmard gate infidelities of HQC, HQC+DRAG, and HQC+OP with different settings of decoherence and leakage parameters in (c) and (d).}\label{single}
\end{figure}

\section{Holonomic quantum gate via dynamical invariant}

To construct holonomic gates, we choose an eigenvector $\left|\mu_{1}(t)\right\rangle$ of dynamical invariant as auxiliary state, which satisfies the following conditions~\cite{Liu2019} of (i) the cyclic evolution $\Pi_{1}(0)=\Pi_{1}(\tau)=|\Phi_{b}\rangle\langle \Phi_{b}|$ with $\chi(0)=\chi(\tau)=0 (2\pi)$, (ii) the von Neumann equation $\frac{d}{d t} \Pi_{k}(t)=-i\left[H_{1}(t), \Pi_{k}(t)\right]$,  where $\Pi_{k}(t)\equiv\left|\mu_{k}(t)\right\rangle\left\langle\mu_{k}(t)\right|$ denotes the projector of the auxiliary basis, and (iii) the elimination of dynamical phase $\gamma_{d}=\int_{0}^{\tau}\langle\mu_{1}(t)|H_{e}(t)|\mu_{1}(t)\rangle \ dt=0$.

Now, we demonstrate how to build up universal arbitrary holonomic single-qubit gates. Here, we choose the auxiliary parameters $\chi (t)=\frac{\pi}{2}\left[1-\cos(\frac{2\pi t}{\tau})\right]$ for a cyclic evolution. After the evolution, the state $|\mu_{1}\rangle$ gains a global phase $\gamma$, i.e., $|\mu_{1}(\tau)\rangle=e^{-i\gamma}|\mu_{1}(0)\rangle$ including both geometric phase $\gamma_{g}$ and dynamical phase $\gamma_{d}$~\cite{Liu2019}. The pure geometric phases $\gamma=\eta_{g}$ can be obtained by using the spin-echo technique by setting $\eta(t)=-\frac{2\pi}{5}\sin(\frac{\pi t}{\tau})\cos(\frac{\pi t}{\tau})-\frac{\pi}{2}$ and $\eta(t)=\frac{2\pi}{5}\sin(\frac{\pi t}{\tau})\cos(\frac{\pi t}{\tau})+\frac{\pi}{2}-\eta_{g}$ for two intervals, $(0, \tau / 2)$ and $(\tau / 2, \tau)$ which erase the accumulated dynamical phase, as shown in Fig. \ref{setup}(c). As a result, the final time evolution operator on the subspace $\{|\Phi_{b}\rangle,|\Phi_{d}\rangle\}$ is given by $U(\tau)=e^{i \gamma}|\Phi_{b}\rangle\langle \Phi_{b}|+| \Phi_{d}\rangle\langle \Phi_{d}|$. Consequently,  the holonomic gate can be spanned into the basis $\{|0\rangle, |1\rangle\}$ as,
\begin{equation}\label{U1}
\begin{aligned} U(\theta, \phi, \gamma) &=\left(\begin{array}{cc}c_{\frac{\gamma}{2}}-i s_{\frac{\gamma}{2}} c_{\theta} & -i s_{ \frac{\gamma}{2}} s_{\theta} e^{i \phi}  \\ -i s_{ \frac{\gamma}{2}} s_{\theta} e^{-i \phi} & c_{\frac{\gamma}{2}}+i s_{\frac{\gamma}{2}} c_{\theta} \end{array}\right) \\ &=\exp \left(-i \frac{\gamma}{2} \mathbf{n} \cdot \sigma\right) \end{aligned} \ ,
\end{equation}
This operation corresponds to a rotation around the axis $\textbf{n}=(\sin{\theta}\cos{\phi},\sin{\theta}\sin{\phi},\cos{\theta})$ by an angle of $\gamma$, which picks up a global phase $\gamma/2$. Therefore, it is feasible to implement the holonomic gate by suitably designing particular $\phi$ and $\gamma$.

When taking the decoherence effect into consideration, the performance of a holonomic gate in Eq.~(\ref{U1}), induced by the Hamiltonian in Eq.  (\ref{INter}), can be evaluated by using a master equation in the Lindblad form~\cite{Lindblad} as,
\begin{eqnarray}  \label{me}
\dot\rho(t) &=& i[\rho(t), H_I(t)]  + \frac{1}{2}\left[ \Gamma_1 \mathcal{L}(\lambda^{+}) + \Gamma_2 \mathcal{L}(\lambda^{z})\right],
\end{eqnarray}
where $\rho(t)$  is the density matrix of the considered system and $\mathcal{L}(A)=2A\rho_1 A^\dagger-A^\dagger A \rho_1 -\rho_1 A^\dagger A$ is the Lindbladian of the operator $A$, and $\lambda^{+}=|0\rangle\langle e|+\sqrt{2}|e\rangle\langle 1|+\sqrt{3}|1\rangle\langle h|$,  and $\lambda^{z}=|e\rangle\langle e|+2|1\rangle\langle 1|+3|h\rangle\langle h|$. In addition, $\Gamma_1^j$  and $\Gamma_2^j$ are the decoherence parameters of the $\{|0\rangle, |1\rangle, |e\rangle, |h\rangle \}$ four-level systems. In our simulation, we have used the following experimental parameters~\cite{Barends2014,chen2016}. The anharmonicity, gate time and decoherence parameter are set as $\alpha=2\pi \times 225$ MHz, $\tau=30$ ns, and
$\Gamma_1=\Gamma_2 \approx 2\pi \times 4$ kHz, respectively. We have investigated the state and gate fidelity of the Z with $\theta=0,\phi=0,\gamma=\pi$ and Hardmard gates with $\theta=\pi/4,\phi=0,\gamma=\pi$ for the initial states $|\psi(0)\rangle=(|0\rangle+|1\rangle)/\sqrt{2}$ and $|\psi(0)\rangle=|0\rangle$. The time-dependence of the state population and the state fidelity dynamics of the Z and Hardmard gates are depicted in Fig.~\ref{Zgate}(a) and~\ref{single}(a), where the state fidelity is defined by $F_{s}=\left|\left\langle\psi_{I} | \psi(\tau)\right\rangle\right|^{2}$. The state fidelities $F_{s}$ of the Z and Hardmard gates are obtained to be $94.36\%$ and $96.86\%$ with the ideal state $|\psi_{I}\rangle=\frac{1}{\sqrt{2}}(|0\rangle-|1\rangle)$ and $|\psi_{I}\rangle=\frac{1}{\sqrt{2}}(|0\rangle+|1\rangle)$. Furthermore, we have also investigated the gate fidelities $F_{g}$ of the Z and Hardmard gates for initial states of the form $|\psi\rangle=\cos\Theta|0\rangle+\sin\Theta|1\rangle$, where a total of 1001 different values of $\Theta$ were uniformly chosen in the range of $[0, 2\pi]$, as shown in Fig. \ref{single}(c). The gate fidelities $F_{H}$ of the Z and Hardmard gates are given by $97.12\%$ and $97.84\%$ with the experimental parameters. Note that through our numerical analysis, we find that the higher-order leakage terms than $|h\rangle$ only produces about 0.01\% gate infidelity. Therefore, the leakage of populations to states higher than $|h\rangle$ can be reasonably ignored.

To analyze the major sources of error, we firstly consider the decoherence effect by setting $\Gamma_{1,2}=0$. Comparing Fig.~\ref{Zgate}(c)~[\ref{single}(c)] with Fig.~\ref{Zgate}(d)~[\ref{single}(d)], we found that the decoherence effect has very little effect on the low fidelity.  Next, to investigate the effect of leakage error, we set $H_{\text{leak}}=0$ but with the consideration of the decoherence effect. And we obtain the gate fidelities of the Z and Hardmard gates as high as $99.93\%$ and $99.92\%$.  Therefore, we are able to construct error budgets for the Z and Hardmard gates with the leakage error $(97.6\%, 96.3\%)$ and decoherence error $(2.4\%, 3.7\%)$ by using the contribution to the total error in percentage.

\section{Elimination of Leakage via Optimal Control}

To reduce the leakage error, we take two methods of the derivative removal by adiabatic gate (DRAG)~\cite{chen2016,Motzoi,Gambetta,Werninghaus2020} and the optimal control technology, called HQC+DRAG and HQC+OP. The goal of our optimization is to implement holonomic gates $U_{\text{hol}}=U(\theta, \phi, \gamma)$ contained within the computational states $\{|0\rangle,|1\rangle\}$. Here, we would like to adjust the pulse parameters of amplitudes $\Omega_{1,2}(t)$ and phases $\phi_{0,1}(t)$ in Eq. (\ref{Original}) to realize the target evolution as~\cite{Motzoi,Gambetta}
\begin{equation}\label{target}
 U_{\text {target}}(\tau)=\mathcal{T} e^{-i \int_{0}^{\tau} H_{I}(t) d t}=e^{i \delta} U_{\mathrm{hol}}(\tau) \oplus U_{\text {out }}(\tau) ,
\end{equation}
where $\mathcal{T}$ is the time ordering operator and $ U_{\text {out }}(\tau)$ is an evolution operator outside of the computational states, $\delta$ denotes a relative phase.

\subsection{HQC+DRAG}
In this subsection, we demonstrate that our scheme is compatible with DRAG to further enhance the gate fidelities of the Z and Hardmard gates against the leakage error. For the optimization of Z gate,
we only need to consider the optimization of one microwave pulse due to $\Omega_{0}=0$. Thus, the correcting pulse of DRAG~\cite{chen2016,Motzoi,Gambetta,Werninghaus2020} is given by
\begin{equation}\label{DRAGZgate}
\begin{aligned}
\tilde{\Omega}_{1}&=[(X_{1}+v_{1} \dot{X_{1}} / 2 \alpha)^{2}+(Y_{1}-v_{2} \dot{Y_{1}} / 2 \alpha)^{2}]^{-1/2} \\
\tilde{\phi}_{1}&=\tan ^{-1}[(Y_{1}-v_{2} \dot{Y_{1}} / 2 \alpha)/(X_{1}+v_{1} \dot{X_{1}} / 2 \alpha)] \ ,
\end{aligned}
\end{equation}
where $X_{i}\equiv\Omega_{i}(t)\cos\phi_{i}(t)$ and $Y_{i}\equiv\Omega_{i}(t)\sin\phi_{i}(t)$. $v_{1,2}$ are the time-independent weighting parameters. In fact, the unwanted transitions of $|0\rangle \leftrightarrow |e\rangle$ and $|1\rangle \leftrightarrow |h\rangle$ of Z gate can be suppressed via numerical optimization of $v_{1,2}$ according to the DRAG correction~\cite{Gambetta}. However, the phase errors due to ac Stark shifting of the above unwanted transitions~\cite{chen2016} will smear the geometric phases $\gamma$ and thus reduce the gate fidelity of Z gate. Here, we modify the phase parameter $\eta_{g}$ for suppressing phase errors. To simultaneously suppress both errors, we can shape the pulses by the numerical optimization of the gate fidelity $F_{g}$ as functions of the weighting parameters $v_{1}$, $v_{2}$ and the phase parameter $\eta_{g}$. After the optimization, we find the gate fidelity $F_{g}$ of Z gate can be significantly improved from $97.12\%$ to $99.76\%$ with the parameters $v_{1}=0$, $v_{2}=0.1$ and $\eta_{g}=1.15\pi$, as shown in Fig.~\ref{Zgate}(c) and~\ref{Zgate}(d).

\begin{figure}[htp]
\centering
\includegraphics[width=8.7cm]{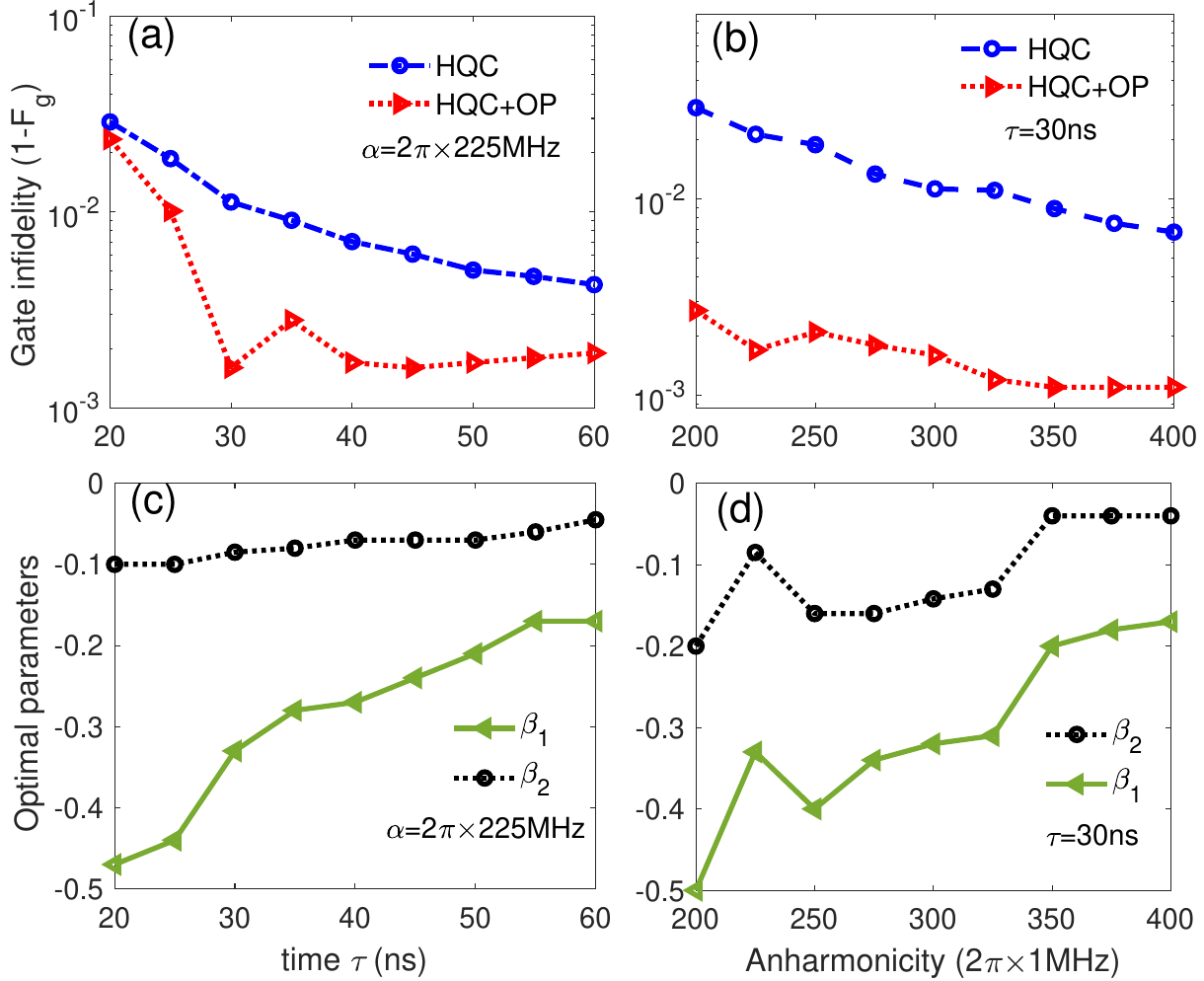}
\caption{ The Hardmard performances of HQC and HQC+OP. The gate infidelities as functions of time $\tau$ (a) and anharmonicity $\alpha$ (b) with the consideration of the decoherence effect.  The optimal parameters $\beta_{1,2}$ vary with time $\tau$ (c) and anharmonicity $\alpha$ (d).}\label{Optimal}
\end{figure}

Besides the above optimizations, we need to optimize the additional parameters of $\Omega_{0}$ and $\phi_{0}$ for Hardmard gate. Using the DRAG gives
\begin{equation}\label{DRAGHgate}
\begin{aligned}
\tilde{\Omega}_{0}&=[(X_{0}+v_{3} \dot{X_{0}} / 2 \alpha)^{2}+(Y_{0}-v_{4} \dot{Y_{0}} / 2 \alpha)^{2}]^{-1/2} \\
\tilde{\phi}_{0}&=\tan ^{-1}[(Y_{0}-v_{4} \dot{Y_{0}} / 2 \alpha)/(X_{0}+v_{3} \dot{X_{0}} / 2 \alpha)] \ ,
\end{aligned}
\end{equation}
where $v_{3,4}$ are also the time-independent weighting parameters. Similar to the optimization of Z gate, we find the gate fidelity $F_{g}$ of Hardmard gate is significantly improved from $97.84\%$ to $99.69\%$ with the parameters $v_{1}=0$, $v_{2}=1.9$, $v_{3}=-5.5$, $v_{4}=5.5$ and $\eta_{g}=1.06\pi$, as shown in Fig.~\ref{single}(c) and~\ref{single}(d).

\begin{figure*}[htbp]
\centering
\includegraphics[width=14cm]{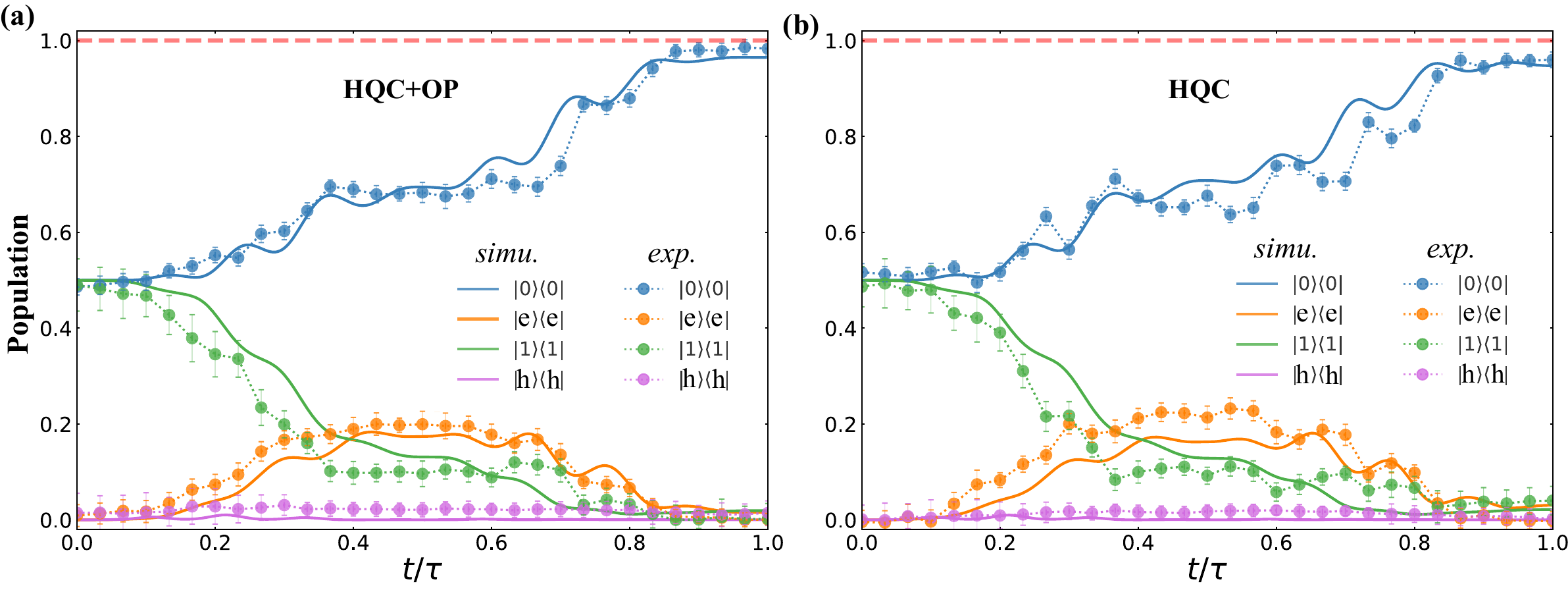}
\caption{ State population and fidelity dynamics of (a) optimal holonomic (HQC-OP) and (b) holonomic (HQC) Hadamard gate, as a function of $t/\tau$ with the initial state being $|\psi(0)\rangle=(|0\rangle+|1\rangle)/\sqrt{2}$. Dots and lines are experimental data and numerical simulation, respectively.}\label{ExP}
\end{figure*}

\subsection{HQC+OP}

As inspired by the DRAG method~\cite{chen2016,Motzoi,Gambetta,Werninghaus2020}, we add two different quadrature components to $\Omega_{0,1}(t)$ with two time-independent weighting parameters $\beta_{1}$ and $\beta_{2}$. With this setting, the correcting pulses of HQC+OP are given by
\begin{equation}
\widetilde{\Omega}_{0}(t)=\left(1+i \beta_{1}\right) \Omega_{0}(t),\quad \widetilde{\Omega}_{1}(t)=\left(1+i \beta_{2}\right) \Omega_{1}(t) \ ,
\end{equation}
where the weighting parameters $\beta_{1,2}$ are also determined by the numerical optimization.

For the optimization of Z gate, we find that the state and gate fidelity can be enhanced from $F_{s}=94.36\%$ and $F_{g}=97.12\%$ to $F_{s}=99.85\%$ and $F_{g}=99.73\%$ with the parameters $\beta_{1}=0$, $\beta_{2}=-0.130$, and $\eta_{g}=0.07\pi$, as shown in Fig.~\ref{Zgate}(b) and~\ref{Zgate}(d). To optimize the Hardmard gate, we numerically choose the parameters of $\beta_{1}=-0.33$, $\beta_{2}=-0.085$ and $\eta_{g}=0.983\pi$ to reduce the leakage and phase error.  With the optimal parameters, the state and gate fidelity of Hadamard gate can be improved from $F_{s}=96.86\%$ and $F_{g}=97.84\%$ to $F_{s}=99.85\%$ and $F_{g}=99.84\%$, respectively, as shown in Fig.~\ref{single}(b-d).

To further illustrate the practicality of our optimized control method, we plot the gate fidelities as functions of the gate time $\tau$ and anharmonicity $\alpha$ of the Xmon qubit with considering decoherence effects. In the Fig. \ref{Optimal}(a) and Fig. \ref{Optimal}(b), we can find that the gate fidelities of the holonomic Hadamard gate can always be improved by choosing appropriate parameters. The optimal parameters $\beta_{1}$, $\beta_{2}$ and $\eta_{g}$ are given in the Table \ref{table1} and shown in the Fig.~\ref{Optimal}(c) and \ref{Optimal}(d). From the Fig. \ref{Optimal}(a), we find a short gate time $\tau=30$ ns with optimal parameters $\beta_{1,2}$, which balance leakage error and the decoherence effect. Consequently, the proposed nonadiabatic holonomic gate in our method can significantly avoid losses with the short gate time.

In general, both HQC+OP and HQC+DRAG can improve the gate fidelities of the holonomic gates. However, the scheme of HQC+OP needs to optimize fewer parameters than HQC+DRAG. Therefore, HQC+OP is easier to experimently implement with fewer resources.

\begin{table}[htb]
\centering \caption{Control parameters of optimal pulses $\beta_{1,2}$ and $\eta_{g}$ for Hardmard gate vary with gate time $\tau$ and anharmonicity $\alpha$.}
\begin{tabular}{cccccccccc}
\hline
 $\tau$  & 20 & 25 & 30 & 35 & 40 & 45 & 50 & 55 & 60\\
   \hline
 -$\beta_{1}$ & 0.47 & 0.44 & 0.33 & 0.28 & 0.27 & 0.24 & 0.21 & 0.17& 0.17 \\
-$\beta_{2}$  & 0.10 & 0.10 & 0.085 & 0.08 & 0.07 & 0.07 & 0.07 & 0.06 & 0.045 \\
 $\eta_{g}/\pi$ & 0.96 & 0.97 & 0.983 & 0.984 & 0.985 & 0.985 & 0.9855 & 0.986 & 0.987 \\
 \hline
  $\alpha/2\pi$  & 200 & 225 & 250 & 275 & 300 & 325 & 350 & 375 & 400 \\
   \hline
 -$\beta_{1}$ & 0.5 & 0.33 & 0.40 & 0.34 & 0.32 & 0.31 & 0.2 & 0.180 & 0.17 \\
-$\beta_{2}$  & 0.2 & 0.085 & 0.016 & 0.016 & 0.014 & 0.013& 0.04& 0.04 & 0.04 \\
 $\eta_{g}/\pi$ & 0.93 & 0.983 & 0.93 & 0.93 & 0.94 & 0.942 & 1.00 & 1.00 & 1.00 \\
 \hline
\end{tabular}\label{table1}
\end{table}

\section{Experimental demonstration of gate performance }

In this section, we experimentally demonstrate the state population and fidelity of HQC and HQC+OP in a supercondcuting Xmon qubit. The four-energy levels of our Xmon qubit are characterized by, $\omega_{0e}/2\pi=4.7114$ GHz, $\omega_{e1}/2\pi=4.4336$ GHz, and $\omega_{1h}/2\pi =4.1146$ GHz. The anharmonicity of the third level in our system is about $\alpha/2\pi\approx0.277$ GHz. The coherence times of our system are $T_1^{0e}=11.37$ $\mu$s, $T_1^{e1}=8.82$ $\mu$s, $T_1^{1h}=7.66$ $\mu$s, $T_2^{0e}=0.87$ $\mu$s, $T_2^{e1}=0.49$ $\mu$s, and $T_2^{1h}=0.20$ $\mu$s. Level spacing of the Xmon qubit can be fine tuned by a bias current on the Z control line. The control microwave pulses are applied to the qubit through the XY control line. The Xmon qubit is coupled to a $\lambda/4 $ resonator ($\omega_{r}/2\pi=6.8298$ GHz) for qubit readout, which is in turn coupled to a transmission line. The state of the Xmon qubit can be deduced by measuring the transmission coefficient $S21$ of the transmission line using the dispersive readout scheme~\cite{Yan2019}.

In our experiment, we take the evolution time of Hadamard gate as $\tau=30$ ns. The state of Xmon qubit is initialized to the superposition state $|\psi(0)\rangle=(|0\rangle+|1\rangle)/\sqrt{2}$. After a Hadamard gate, the ideal final state should be $|\psi(\tau)\rangle=|0\rangle$. The time-dependence of the state populations of the optimal holonomic and holonomic Hadamard gates are depicted in Fig. \ref{ExP}(a) and \ref{ExP}(b). Here, the experimental state fidelities $F_{s}$ with (without) the consideration of optimal control are obtained to be 98.2\% (96.0\%), where the optimal parameters are taken as $\beta_{1}=-0.24$, $\beta_{2}=-0.055$ and $\eta_{g}=\pi$. We obtain numerical Hardmard fidelities with (without) the consideration of optimal control as 98.28\%  (97.3\%) by using a master-equation scheme, which take the decoherence into account. The results are in good agreement with the experimental results.

\section{conclusion}

In summary, we have demonstrated a novel dynamical-invariant based nonadiabatic holonomic gate scheme in a weak anharmonic superconducting Xmon qubit. Moreover, the proposed scheme is compatible with the optimal control technology, which substantially improves the gate fidelities by minimizing the leakage error. Numerical simulation and experimental results show that the leakage rates can be significantly reduced by choosing suitable control parameters. Our HQC scheme provides a promising way towards fault-tolerant quantum computation in a realistic solid-state system.

\acknowledgments
This work is supported by the Key-Area Research and Development Program of Guangdong Province (Grant No.2018B030326001), the National Natural Science Foundation of China (Grant No.11875160 and No.11874156), the Natural Science Foundation of Guangdong Province (Grant No.2017B030308003), the National Key R\& D Program of China (Grant No.2016YFA0301803, the Guangdong Innovative and Entrepreneurial Research Team Program (Grant No.2016ZT06D348), the Economy, Trade and Information Commission of Shenzhen Municipality (Grant No.201901161512), the Science, Technology and Innovation Commission of Shenzhen Municipality (Grant No. JCYJ20170412152620376, No. JCYJ20170817105046702, and No. KYTDPT20181011104202253). We particularly thank Prof. Yuanzhen Chen and Dr. Jingjing Niu at Southern University of Science and Technology, where all the experimental data were taken, for providing access to the experimental facilities.
We also thank Prof. S.-L. Su and Dr. X.-M. Zhang for valuable discussions.

\begin{appendix}

\end{appendix}

\end{document}